\documentclass[a4paper,11pt]{elsarticle}
\pdfoutput=1 % if your are submitting a pdflatex (i.e. if you have
\DeclareUnicodeCharacter{2212}{-}
\usepackage[T1]{fontenc} % if needed
\usepackage{graphicx}
\usepackage{subfigure}
\graphicspath{{img/}}
\usepackage{amsmath}
\usepackage{amssymb}
\usepackage{amsfonts}
\usepackage{mathrsfs}
\usepackage{lscape}
\usepackage{hyperref}
\usepackage{verbatim}
\usepackage{xcolor}
\usepackage{bm}
\usepackage{tabularx}
\usepackage{mathtools}
\usepackage{ulem}
\usepackage{float}
\newcommand{\be}{\begin{equation}}
\newcommand{\ee}{\end{equation}}
\newcommand{\ba}{\begin{eqnarray}}
\newcommand{\ea}{\end{eqnarray}}

\newcommand{\pide}{\pi_{\rm de}}
\renewcommand{\H}{{\cal H}}

\newcommand{\epi}{e_\pi}

\newcommand{\ceff}{c_{\rm eff}}
\def\bea{\begin{eqnarray}}
\def\eea{\end{eqnarray}}
\def\eqi{\begin{equation}}
\def\eqf{\end{equation}}
\def\eqia{\begin{eqnarray}}
\def\eqfa{\end{eqnarray}}
 
\def\Mpl2{M_{\rm{pl}}^2}

\definecolor{darkgreen}{rgb}{0,0.6,0}
\begin{document}

\begin{frontmatter}

\title{Lensing convergence and anisotropic dark energy in galaxy redshift surveys}

\author{Wilmar Cardona}
\ead{wilmarcardonac@gmail.com}
\address{Departamento de F\'isica, Universidad Antonio Nari\~no, Cra $3$ Este No. $47$A-$15$, Bogot\'a DC, Colombia}
\address{Departamento de F\'isica, Universidad del Valle, Ciudad Universitaria Mel\'endez, 760032, Cali, Colombia}

\author{Rub\'{e}n Arjona}
\ead{ruben.arjoname@gmail.com}

\author{Savvas Nesseris}
\ead{savvas.nesseris@csic.es}
\address{Instituto de F\'isica Te\'orica UAM-CSIC, Universidad Autonoma de Madrid, Cantoblanco, 28049 Madrid, Spain}

\begin{abstract}
Analyses of upcoming galaxy surveys will require careful modelling of relevant observables such as the power spectrum of galaxy counts in harmonic space $C_\ell(z,z')$. We investigate the impact of disregarding relevant relativistic effects by considering a model of dark energy including constant sound speed $c_{\rm eff}^2$, constant equation of state $w$, and anisotropic stress sourced by matter perturbations $\pi$. Cosmological constraints were computed using cosmic microwave background anisotropies, baryon acoustic oscillations, supernovae type Ia, and redshift space distortions. Our results are consistent with $w=-1$, $c_{\rm eff}^2=1$, and $\pi=0$. Then, a forecast for the performance of an Euclid-like galaxy survey was carried out also adding information from other probes. Here we show that, regardless of the galaxy survey configuration, neglecting the effect of lensing convergence will lead to substantial shifts in the galaxy bias $b_0$ and the neutrino mass $\sum m_\nu$. Shifts in the dark energy sound speed and anisotropic stress also appear, but they depend on the survey configuration and hence lack robustness. While neglecting lensing convergence also leads to a Hubble constant $H_0$ moving downwards, the significance of the shift is not big enough to play a relevant part in the current $H_0$ tension.
\end{abstract}


\end{frontmatter}

%\date{\today}

%\maketitle
%\flushbottom

%%%%%%%%%%%%%%%%%%%%%%%%%%%%%%%%%%%%%%%%%%%%
\section{Introduction}
\label{sec:intro}
%%%%%%%%%%%%%%%%%%%%%%%%%%%%%%%%%%%%%%%%%%%%

More than two decades after the discovery of the accelerating expansion of the Universe~\cite{Riess:1998cb,Perlmutter:1998np} there is not yet a convincing explanation for this phenomenon. Although the $\Lambda$CDM model is in very good agreement with most of the current data sets~\cite{Abbott:2017wau,Aghanim:2018eyx,PhysRevLett.122.171301}, the standard model of cosmology is not the only successful phenomenological fit to the data. 
%As a result two leading alternative approaches have emerged: on the one hand, dynamical Dark Energy (DE) models, and on the other hand, the so--called Modified Gravity (MG) models. 
As a result, two complementary approaches have emerged: dynamical Dark Energy (DE) and  Modified Gravity (MG) models. 
DE models may utilize scalar fields (e.g., quintessence, K--essence, phantom) as a new ingredient in the model thus providing the pressure conditions which accelerate the Universe at late times~\cite{Copeland:2006wr,Ratra:1987rm}. MG models change the gravity sector in the Einstein equations (e.g., $f(R)$ models, massive gravity, DGP) in order to achieve the recent speeding--up phase in the Universe~\cite{Clifton:2011jh}.

A key goal for current and forthcoming experiments such as \textit{Euclid} \cite{EUCLID:2011zbd}, DESI \cite{DESI:2016fyo}, LSST \cite{LSSTScience:2009jmu}, LiteBIRD \cite{Hazumi:2019lys}, LIGO \cite{LIGOScientific:2014pky} or LISA \cite{LISA:2017pwj}, 
%(e.g., galaxy surveys, CMB experiments, gravitational wave detectors) 
will be to discriminate among different explanations for the current accelerated expansion. Both DE models and MG models imply modifications to the Friedmann equations. %However, these modifications can be negligible with respect to the standard model: in other words, the models can be fully degenerate at the background level. 
However, these modifications can give the same equation of state parameter $w$ at the background level, thus certain DE and MG models can be degenerate. A clear example of this situation is the so--called `designer model' where it is possible -- in an effective fluid interpretation of modifications to General Relativity -- to find a family of $f(R)$ models having equation of state $w=-1$~\cite{Nojiri:2006ri,Capozziello:2006dj,Pogosian:2007sw,Multamaki:2005zs,delaCruzDombriz:2006fj,Nesseris:2013fca,PhysRevD.99.043516}. It has been shown that it is also possible to find `designer models' in the context of Horndeski and scalar-vector-tensor theories (see, for instance, Refs.~\cite{Arjona:2019rfn,Cardona:2022lcz}).

It is thus necessary to go beyond the background level in order to break degeneracies among different models. Perturbations in both Cosmic Microwave Background (CMB) and matter distribution are decisive observables because their statistical properties allow us to further distinguish cosmological models~\cite{Eisenstein:1998hr,Lewis:2002ah,Tegmark:2003ud}. Although invisible, DE and Dark Matter (DM) perturbations have an impact on these observables which can in turn be used to constrain DE and MG models~\cite{Cardona:2014iba,PhysRevD.85.123529,Amendola:2007rr,Sapone:2009mb,dePutter:2010vy} .

Here we consider an imperfect fluid parametrised by its equation of state $w$, its sound speed $c_{\rm{s}}^2$, and its anisotropic stress $\pi_{\rm fld}$~\cite{Kunz:2012aw}, but as we describe the DE fluid in its rest frame, we can neglect the heat flux \cite{Ballesteros:2012kv}. %For DE models both $c_{\rm s}^2$ and $\pi_{\rm fld}$ are currently undetermined and their detection might be very significant. 
Currently, both $c_{\rm s}^2$ and $\pi_{\rm fld}$ are not well-constrained by observational data, with data showing a consistency with $\Lambda$CDM \cite{Arjona:2018jhh}, however any deviations from constant values could provide hints to beyond the vanilla $\Lambda$CDM scenario. This is the case as the sound speed $c_{\rm s}^2$ is related to the level of clustering in DE perturbations~\cite{dePutter:2010vy}, while the anisotropic stress can act as a source for matter perturbations and therefore might leave detectable traces in the angular matter power spectrum~\cite{Cardona:2014iba,Koivisto:2005mm,Mota:2007sz,Garcia-Arroyo:2020iou}. The presence of anisotropic stress might be conclusive to discriminate MG and DE models~\cite{Saltas:2010tt,Kunz:2006ca, Sobral-Blanco:2021cks}.

DE anisotropic stress $\pide$ plays a particularly important part in distinguishing models for the late--time universe. On the one hand, simple single--field DE models, such as quintessence and K--esence, have a vanishing anisotropic stress $\pide=0$~\cite{Kunz:2012aw}. On the other hand, MG models, such as $f(R)$ models and DGP, generically possess a non--zero anisotropic stress~\cite{Pogosian:2007sw,Koyama:2005kd,Tsujikawa:2007gd,PhysRevD.99.043516,Arjona:2019rfn,Cardona:2022lcz}. Therefore the detection of a non vanishing DE anisotropic stress would rule out the simplest DE models thus throwing light on the problem of the late--time accelerating universe~\cite{Saltas:2010tt,Kunz:2006ca}.

Signatures of DE anisotropic stress, which are also present in MG theories, are expected from differences in the gravitational potentials $\psi \neq \phi$ as well as their late--time evolution, that is, $\phi',\, \psi' \neq 0$~\cite{Amendola:2007rr}. The Integrated Sachs Wolf (ISW) effect and the lensing potential are key when trying to get information about the gravitational potentials and their evolution \cite{Bertschinger:2011kk}. Two main complementary probes are known to be sensitive to these effects: fluctuations in both CMB and Number Counts (NC). Therefore, by measuring fluctuations in the CMB and NC we would in principle be able to constrain important quantities such as the neutrino mass scale and DE anisotropic stress. Achieving this goal is of crucial importance for understanding the cosmological evolution and will require careful modelling of the underlying physical phenomena whether biased constraints are to be avoided.

A couple of phenomenological models for DE anisotropic stress were proposed in Ref.~\cite{Cardona:2014iba}. Firstly, it was considered that the DE anisotropic stress is sourced by the matter comoving density perturbation, namely, $\pide \propto \Delta_{\rm m}$, which is the kind of anisotropic stress present in DGP models and possibly in interacting DE models. Secondly, the authors considered a model where the DE anisotropic stress is internally sourced by DE perturbations, that is, $\pide \propto \Delta_{\rm de}$. This sort of anisotropic stress is generically found, for instance, in MG models when modifications to gravity are interpreted as an effective fluid~\cite{PhysRevD.99.043516,Arjona:2019rfn,Cardona:2022lcz}. 

By using mainly CMB data and background data (e.g., supernova type Ia [SNe], Baryon Acoustic Oscillations [BAO]) constraints  on these kinds of DE anisotropic models were found in Ref.~\cite{Cardona:2014iba}. Recently, the internally sourced DE anisotropic stress was further studied in the context of interacting DE-DM scenarios in Ref.~\cite{Yang:2018ubt}. The results in Refs.~\cite{Cardona:2014iba,Yang:2018ubt} show that externally sourced DE anisotropic stress is consistent with zero and that there is still room for a non--vanishing internally sourced DE anisotropic stress. These constraints could be significantly improved with the inclusion of lensing data as well as NC from upcoming and ongoing galaxy surveys such as Euclid.\footnote{https://www.euclid-ec.org/}

Observations of the matter density fluctuations are important because they provide complementary information to the CMB anisotropies thus allowing the breaking of degeneracies in cosmological parameters as well as tightening constraints. Quantities which are currently unconstrained such as the neutrino mass scale, DE anisotropic stress, and the DE sound speed could in principle be determined by the use of NC in the analysis. As time goes by, galaxy surveys are probing scales comparable to the horizon and careful modeling of NC is required~\cite{Foglieni:2023xca}. Therefore, relativistic effects such as lensing convergence cannot be neglected any longer since this could lead to spurious detections of the neutrino mass~\cite{Cardona:2016qxn} or of non-gaussianities \cite{Grimm:2020ays}. This would be important: the neutrino mass enters the background density and affects the expansion of the Universe, especially at early time; suppression of matter power spectrum on small scales is also related to the neutrino mass scale~\cite{PhysRevLett.95.011302}. For exact analytic expressions for the density and pressure of massive neutrinos, see Ref.~\cite{arjona2019analytic}.

The importance of the inclusion of lensing convergence in the analysis of future galaxy surveys has been previously considered. The impact on the detection of local-type non-Gaussianity was studied in Refs.~\cite{Namikawa:2011yr,Camera:2014sba,Camera:2018jys}  where authors concluded that a proper account of the magnification effect is quite essential for an unbiased estimate of primordial non-Gaussianity. In Ref.~\cite{Duncan:2013haa} the effect of neglecting lensing convergence when constraining cosmological parameters such as the equation of state $w$ was investigated; the study revealed that if the flux magnification is incorrectly neglected, then bias in inferred $w$ can be many times larger than statistical uncertainties for a Stage IV space-based photometric survey. Authors in Ref.~\cite{Cardona:2016qxn} showed that no inclusion of lensing convergence would lead to biased cosmological constraints including a spurious detection of the neutrino mass scale. Similar conclusions were found in Ref.~\cite{Cardona:2022mdq} for a model featuring a momentum transfer between dark matter and dark energy.  The importance of taking into consideration lensing magnification was also studied in Ref.~\cite{PhysRevD.97.023537,Villa:2017yfg} where authors considered extensions to the $\Lambda$CDM model, Horndeski-like parametrisations of scalar-tensor theories, and a large-scale contribution of primordial non-Gaussianity to the galaxy power spectrum. They confirmed that it will be necessary to model and account for lensing magnification in order to avoid strong biases on dark energy parameters and the sum of neutrino masses. Recently in Ref.~\cite{Euclid:2021rez} the authors carefully inquired into the impact lensing magnification might have on analyses of galaxy clustering by using realistic specifications for the local count slope based on the Euclid Flagship simulation. It was shown that the impact of including/neglecting magnification in the analysis depends on measurements of the local count slope.

Previous studies have shown the importance of including lensing convergence in analyses of upcoming galaxy surveys for a number of cosmological models. However, the impact of neglecting this contribution to the galaxy power spectrum has thus far not been studied for cosmological models including both DE sound speed and DE anisotropic stress. Here we will address that problem. Firstly, using available data sets we compute cosmological constraints on a constant DE sound speed as well as an externally sourced DE anisotropic stress. Secondly, we use our baseline constraint to feed a forecast for the performance of an Euclid-like galaxy survey. We determine whether new information from NC could pin down $\pide$ and $c_{\rm s}^2$. Thirdly, we examine to what extent cosmological models taking into consideration DE with $w$, $\pide$, $c_{\rm s}^2$ can remove the bias on the neutrino mass and in the equation of state found in Refs.~\cite{Duncan:2013haa,Cardona:2016qxn,PhysRevD.97.023537,Cardona:2022mdq}, if lensing convergence is neglected in the analysis. 

The paper is organized as follows. In Section~\ref{sec:ade} we discuss the perturbation equations for a generic fluid described by equation of state, sound speed, and anisotropic stress; we also give details about the phenomenological DE anisotropic stress model that we consider. Then in Section~\ref{sec:constraints} we present cosmological constraints using available data sets. We explain the way we carry out the forecast in Section~\ref{sec:forecast}. We conclude in Section~\ref{sec:conclusions}.

\section{Anisotropic dark energy}
\label{sec:ade}

Since astrophysical observations indicate that the Universe on large scales is statistically homogeneous and isotropic~\cite{Hogg:2004vw,Ade:2015hxq,Marinoni:2012ba},  we will assume a Friedmann-Lema\^{i}tre-Robertson-Waker (FLRW) metric including tiny inhomogeneities which can be treated within linear perturbation theory. The cosmological model that we will investigate is a relatively simple extension of the standard model $\Lambda$CDM. Throughout the paper we assume flatness, include massive neutrinos with a normal mass hierarchy (dominated by the heaviest neutrino mass eigenstate), and model DE as a fluid described by three quantities, namely: a constant equation of state $w$, a DE fluid rest-frame constant sound speed $\ceff^2$, and anisotropic stress $\pide$. 
 
Below we provide linear order perturbation equations for a generic fluid including anisotropic stress and  also discuss the DE anisotropic stress model that we study.

\subsection{Perturbation equations}

In the longitudinal gauge the perturbed FLRW metric reads
\begin{eqnarray}
ds^2 & \equiv & g_{\mu\nu}dx^\mu dx^\nu \\
  & = & a(\eta)^2 \left\{-[1+2\psi\,(\eta,\textbf{x})] d\eta^2 + [1-2\phi(\eta,\textbf{x}) ] d\vec{x}^2 \right\}, \nonumber
\label{Eq:perturbed-metric}
\end{eqnarray}
where $\eta$ is the conformal time, $a$ is the scale factor, and $\psi$ and $\phi$ are the gravitational potentials.\footnote{In this paper we set the speed of light $c=1$ and adopt the convention that a prime stands for the derivative with respect to the conformal time (e.g., $f(\eta)'\equiv \dfrac{df}{d\eta}$). Also, the semi-colon ``;'' is a covariant derivative.} The conservation of the energy-momentum tensor $T^{\mu\nu}$ for a generic fluid, that is, $T^{\mu\nu}{}_{;\nu}=0$, leads to the continuity and Euler equations
\begin{equation}
\delta' + 3 \H \left( \frac{\delta P}{\rho} - w \delta \right) + (1 + w) k v - 3 (1 + w)\phi' = 0,
\label{Eq:continuity}
\end{equation}
\begin{equation}
v' + \H (1 - 3 c_a^2) v - k \left( \psi + \frac{\delta P}{\rho (1+w)} - \frac{2 \pi_{\rm fld}}{3(1+w)} \right) = 0,
\label{Eq:euler}
\end{equation}
where we define the adiabatic sound speed
\begin{equation}
c_a^2 \equiv w - \frac{w'}{3 \H (1+w)},
\label{Eq:adiabatic-sound-speed}
\end{equation}
the density contrast
\begin{equation}
\delta \equiv \frac{\delta \rho}{\rho},
\label{Eq:density-contrast}
\end{equation}
and the equation of state
\begin{equation}
w \equiv \frac{P}{\rho},
\end{equation}
where all perturbation quantities are in the conformal Newtonian gauge. Here $v$ stands for the velocity perturbation, $k$ is the wave-number, $\delta P$ the pressure perturbation, $\pi_{\rm fld}$ is the anisotropic stress of the fluid, $\rho$ the fluid energy density, $P$ the fluid pressure, and the conformal Hubble parameter is
\begin{equation}
\H \equiv \frac{a'}{a} = a H,
\label{Eq:conformal-Hubble-parameter}
\end{equation}
with $H$ the physical Hubble parameter.

In this paper we will only focus on scalar perturbations. There are consequently two independent Einstein equations which can be written as
\begin{equation}
-k^2 \phi = 4 \pi G a^2 \sum_i \rho_i \Delta_i , 
\label{Eq:Poisson}
\end{equation}
\begin{equation}
k^2 (\phi - \psi) = 8 \pi G a^2 \sum_i \rho_i \pi_i ,
\label{Eq:slip}
\end{equation}
where the index $i$ runs over different matter species (e.g., radiation, neutrinos, baryonic matter, dark matter, dark energy), $G$ is the bare Newton's constant, and  the comoving density perturbation is defined as
\begin{equation}
\Delta_i \equiv \delta_i + 3 ( 1 + w_i ) \H \frac{v_i}{k}.
\label{Eq:comoving-density-perturbation}
\end{equation}
We will follow Ref.~\cite{Cardona:2014iba} and model the pressure perturbation as
\begin{equation}
\frac{\delta P}{\rho} = c_{\rm s}^2 \,\delta + 3(1+w)(c_{\rm s}^2 -  c_a^2)\frac{\H}{k}v,
\label{Eq:pressure-perturbation}
\end{equation}
where we define the effective, non--adiabatic, sound speed of the fluid in its rest--frame as $\partial_\mu P \equiv c_{\rm s}^2 \partial_\mu \rho$, while the second term in Eq.~\eqref{Eq:pressure-perturbation} is a gauge term. It is instructive to rewrite the system of first order differential equations given by Eqs.~\eqref{Eq:continuity}-\eqref{Eq:euler}, as a single second order differential equation. Combining these two equations we obtain
\begin{eqnarray}
\delta'' &+& \left(1-6w\right) \H \delta'
    + 3 \H \left(\!\frac{\delta P}{\rho}\!\right)' - 3 \H w' \delta \nonumber \\
    &+& 3 \Big[(1-3w)\H^2+\H'\Big] \left(\frac{\delta P}{\rho}-w\delta\right) \nonumber \\
    &=& 3(1+w)\left[\phi'' + \left(1-3w +\frac{w'}{(1+w)\H}\right) \H \phi' \right] \nonumber \\
    &-& k^2\left[ (1+w)\psi + \frac{\delta P}{\rho} -\frac{2}{3}\pi_{\rm fld} \right].
\label{Eq:second-order-density-perturbations}
\end{eqnarray}
Under both sub-horizon and quasi-static approximations, the right-hand side of Eq. \eqref{Eq:second-order-density-perturbations} becomes
\be
k^2 \left[ (1+w)\psi + \frac{\delta P}{\rho} -\frac{2}{3}\pi_{\rm fld} \right]
\approx  k^2 \left[  c_{\rm s}^2 \Delta -\frac{2}{3}\pi_{\rm fld} \right] \, ,
\label{Eq:effsound-1}
\ee
where we use Eqs.~\eqref{Eq:comoving-density-perturbation}-\eqref{Eq:pressure-perturbation} and neglect terms $\propto v/k$. Also, at late time the anisotropic stress from radiation or neutrinos can be safely neglected. Therefore it becomes apparent that the anisotropic stress might act as a source for density perturbations which in turn can have an impact in their stability. 

In what follows we will assume that DE can be modeled as a generic fluid with a constant equation of state $w$, DE rest-frame constant sound speed $\ceff^2$, and anisotropic stress $\pide$ given by the model in the next subsection. These assumptions simplify the adiabatic sound speed in Eq.~\eqref{Eq:adiabatic-sound-speed} (it becomes $c_{\rm{a}}^2=w$) as well as the second order differential equation Eq.\eqref{Eq:second-order-density-perturbations} where terms having $w'$ vanish. 

We note that a redefinition of variables describing DE perturbations might be useful in some particular cases. If we define $V\equiv(1+w)kv$, then Eqs.~\eqref{Eq:continuity}-\eqref{Eq:euler} read 
\begin{equation}
\delta' + 3 \H \left( \frac{\delta P}{\rho} - w \delta \right) + V - 3 (1 + w)\phi' = 0,
\label{Eq:continuity-new}
\end{equation}
\begin{equation}
V' + \H (1 - 3 c_a^2) V - k^2 \left( (1+w)\psi + \frac{\delta P}{\rho} - \frac{2 \pi_{\rm fld}}{3} \right) = 0,
\label{Eq:euler-new}
\end{equation}
where $w$ is a constant. Therefore, the system of differential equations remains well defined even when $w=-1$.

\subsection{Externally sourced dark energy anisotropic stress model}

A number of reasons motivate the study of externally sourced DE anisotropic stress. Firstly, when considering the quasi-static limit, it is known that in the Dvali-Gabadadze-Porrati (DGP) cosmological model the gravitational potentials are directly linked to the matter perturbations via  functions which depend on the scale factor \cite{Koyama:2005kd}.  As a result, the difference in the gravitational potentials, $\phi - \psi$, hence the anisotropic stress, also depends on the matter perturbations \cite{Kunz:2006ca}. Secondly, since the nature of the dark sector is largely unknown there exist cosmological models where dark matter and  dark energy are allowed to interact with each other (see, for instance, Ref.~\cite{Amendola:1999er}). Thirdly, note that in fairly general scalar-vector-tensor theories under both sub-horizon and quasi-static approximations the effective dark energy anisotropic stress can also be sourced by dark matter perturbations~\cite{Arjona:2019rfn,PhysRevD.99.043516,Cardona:2022lcz}. 

A simple, plausible DE anisotropic stress model sourced by matter perturbations is \cite{Cardona:2014iba}
\begin{equation}
\pide = \epi \Delta_{\rm m},
\label{Eq:model-1-dea}
\end{equation}
where $\epi$ is a constant, and subscripts `de' and `m' respectively stand for DE and matter. We implemented the model in the Boltzmann code \texttt{CLASS} \cite{Blas:2011rf} which allows us to solve the full system of differential equations (i.e., background and linear order perturbations) and compute observables. In particular, in the case of the linear order perturbations \texttt{CLASS} solves numerically Eqs.~\eqref{Eq:Poisson}-\eqref{Eq:euler-new}. 

\begin{figure*}[h]
\centering
\includegraphics[width=\textwidth]{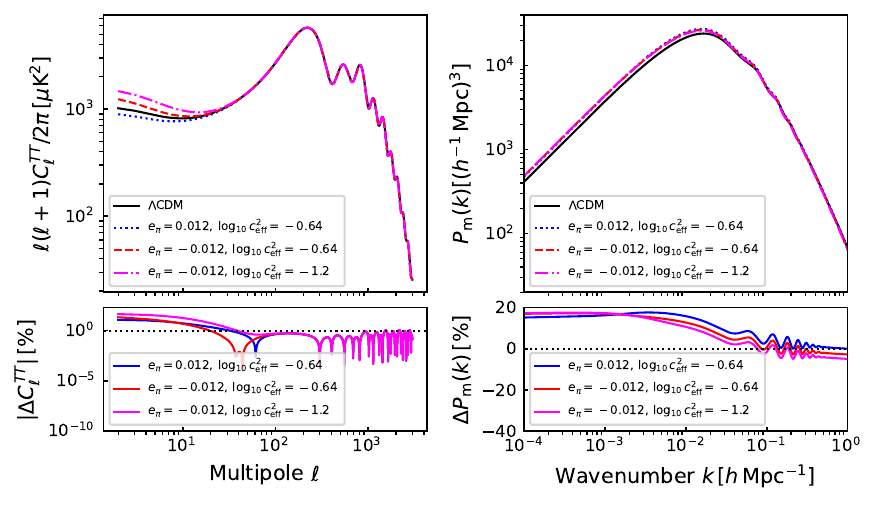}
\caption{Left: CMB TT angular power spectrum accompained by percentage difference relative to $\Lambda$CDM in the lower panel. Right: matter power spectrum at redshift $z=0$ and its percentage difference relative to $\Lambda$CDM in the lower panel. Black solid lines depict the result for the $\Lambda$CDM model using the baseline cosmological parameters reported by the Planck Collaboration~\cite{Aghanim:2018eyx}. Results for the Anisotropic DE (ADE) model were computed setting  cosmological parameters to $\omega_{\rm b}=0.02250$, $\omega_{\rm cdm}=0.1187$, $H_0=69.89\,\mathrm{km}\,\mathrm{s}^{-1}\,\mathrm{Mpc}^{-1}$, $\ln 10^{10}A_{\rm s}=3.039$, $n_{\rm s}=0.9664$, $\tau_{\rm reio}=0.0531$, $\sum m_\nu=0.085\,\mathrm{eV}$, $w=-1.07$.}
\label{Fig:phenomenology}
\end{figure*}

We show a few examples in Fig.~\ref{Fig:phenomenology} for the CMB TT angular power spectrum and the matter power spectrum $P(k,z=0)$. We see  the main effect of a non-vanishing anisotropic stress appears on large scales. 
However, it is interesting to note that while the effect of the anisotropic stress can increase or decrease the power at large scales depending on the sign of $\epi$ (see left panel of Fig.~\ref{Fig:phenomenology}), in the case of the matter power spectrum the situation is different, as regardless of the sign of $\epi$, the power always increases at large scales. The difference for this is that in the former case, the ISW kernel depends on the derivative of the growth, while in the latter the matter power spectrum is the square of the growth itself \cite{Song:2006ej}. On small scales a non-vanishing DE anisotropic stress along with a non trivial DE sound speed might lead to a suppression of power in matter density perturbations, hence possibly alleviating the tension in the strength of matter clustering $\sigma_8$. Note also that on small scales the CMB angular power spectrum is not strongly affected by the presence of relative small DE anisotropic stress. 

\section{Cosmological constraints}
\label{sec:constraints}

In this work we are interested in determining to what extent new information from ongoing and upcoming galaxy surveys will help in pinning down parameters in the ADE model. Nevertheless, in order to fully exploit the power of new experiments and compute tight constraints, we need to add information from available data sets. Firstly, information from the CMB and other probes is complementary and helps breaking possible degeneracies. Secondly, it speeds up the forecasts that we carry out in Section~\ref{sec:model-I}.  

Since the constraints are computed for a high number of parameters (i.e., $36-37$ including nuisance and cosmological parameters), it is convenient to use Markov Chain Monte Carlo (MCMC) techniques~\cite{Lewis:2002ah,2017ARA&A..55..213S}. The procedure is basically as follows. Firstly, we implement the cosmological model in the Boltzmann solver \texttt{CLASS}.\footnote{Modified \texttt{CLASS} code reproducing results in this work can be found in the GitHub branch \texttt{LDE} of the repository  \href{https://github.com/wilmarcardonac/EFCLASS.git}{\texttt{EFCLASS}}.} For a given set of cosmological parameters the code computes background quantities and solves the differential equations governing the linear order perturbations. \texttt{CLASS} is then able to predict the statistical properties for observables such as the CMB angular power spectrum and the matter power spectrum. 

Secondly, we sample the parameter space with the code \texttt{MontePython}~\cite{Audren:2012wb,Brinckmann:2018cvx} that works along with our modified version of \texttt{CLASS}. We choose the default Metropolis-Hastings algorithm in the code and go ahead with the analysis starting from a randomly chosen point within the allowed region.\footnote{For cosmological parameters also analysed by the Planck Collaboration we use a prior range as given in Table 1 of Ref.~\cite{Ade:2013zuv}. Other parameters use the prior range specified in Table~\ref{Table:prior-MGmodels}.} We begin with a diagonal covariance matrix which is updated from time to time until achieving an acceptance rate $\approx 0.25$. Then, the covariance matrix is frozen and the comparison of theoretical predictions against observations is carried out $\sim 10^6$ times so that the Gelman-Rubin statistic $R$ for each parameter converges, where $R$ is defined as $R:=\sqrt{\frac{\mathrm{Var}(\theta)}{W}}$, we have set $\mathrm{Var}(\theta)=\left(1-\frac{1}{n}\right)W+\frac{1}{n} B$, while $\theta$ corresponds to the parameters sampled in the MCMC, $W$ is the variance of a chain, $B$ is the variance of the means between chains, $n$ is the length of the chains, after discarding the points in the burn-in phase \cite{Gelman_Rubin}. Typically, for convergence we require $R-1\lesssim 10^{-2}$ for each parameter (see Ref.~\cite{Gelman_Rubin}).

Furthermore, we sample over all the cosmological parameters in our model: the baryon density today $\omega_{\rm{b}} \equiv \Omega_{\rm{b}} h^2$; the cold dark matter density today $\omega_{\rm{cdm}}\equiv \Omega_{\rm{cdm}}h^2$; $100\times$ angular size of sound horizon at redshift $z_\star$ (redshift for which the optical depth equals unity) $100 \theta_{\star}$; log power of the primordial curvature perturbations $\ln 10^{10}A_{\rm{s}}$; scalar spectrum power-law index $n_{\rm{s}}$; Thomson scattering optical depth due to reionisation $\tau_{\rm reio}$; the sum of neutrino masses $\sum m_\nu(\rm{eV})$; and the DE parameters $w$, $\log_{10} \ceff^2$, $\epi$, however we then marginalize them when we make the relevant $1D$ or $2D$ plots (except of course for the ones shown).

In our analysis we include the following data sets and modify likelihoods already implemented in \texttt{MontePython} when necessary (e.g., when including the Hubble constant). Background parameters are mainly constrained via Pantheon supernovae (\texttt{SNe}) from~\cite{Scolnic:2017caz}, measurements of Baryon Acoustic Oscillations (\texttt{BAO}) from Refs.~\cite{Alam:2016hwk,Beutler:2011hx,Ross:2014qpa}, and the SHOES measurement of the Hubble constant (\texttt{H0}) from Ref.~\cite{Riess:2021jrx} that we take in as a Gaussian prior. Note that the most recent compilation of Cepheids-SNe Ia is provided in the Pantheon+ sample \cite{riess2022comprehensive}, featuring a substantial increase in the number of objects compared to the original Pantheon sample. As for constraining the linear perturbations we utilise a number of probes. Firstly, we incorporate CMB lensing (\texttt{lensing}) as well as temperature and polarisation anisotropies of the CMB (\texttt{TTTEEE}) measured by the Planck Collaboration~\cite{Aghanim:2018eyx}. Secondly, a likelihood for Redshift-Space-Distortions (\texttt{RSD}) including a compilation of measurements that is not implemented in the default version of \texttt{MontePython} (for details see Ref.~\cite{Arjona:2020yum}).          

\begin{table}[http]
\centering
\begin{tabular}{c c}
\hline
      %\multicolumn{2}{c}{Flat prior bounds} \\
\hline
Parameter & Range \\%& text & text & text & text \\
\hline
$w$ & $[-2,-0.3]$\\
%$ \log_{10} c_s^2$ & $[-10,0]$\\
$ \epi $ & $[-\infty,\infty]$\\
$ \log_{10} \ceff^2$ & $[-10,0]$\\
%$\fpi$ & $[0,\infty]$ \\
%$ \log_{10} \gpi$ & $[-5,5]$\\
\hline
\end{tabular}
\caption{MCMC analyses use flat prior distributions shown here. Prior range for other parameters is set as in Table 1 of Ref. \cite{Ade:2013zuv}}
\label{Table:prior-MGmodels}
\end{table}

\begin{figure*}[h]
\centering
\includegraphics[width=\textwidth]{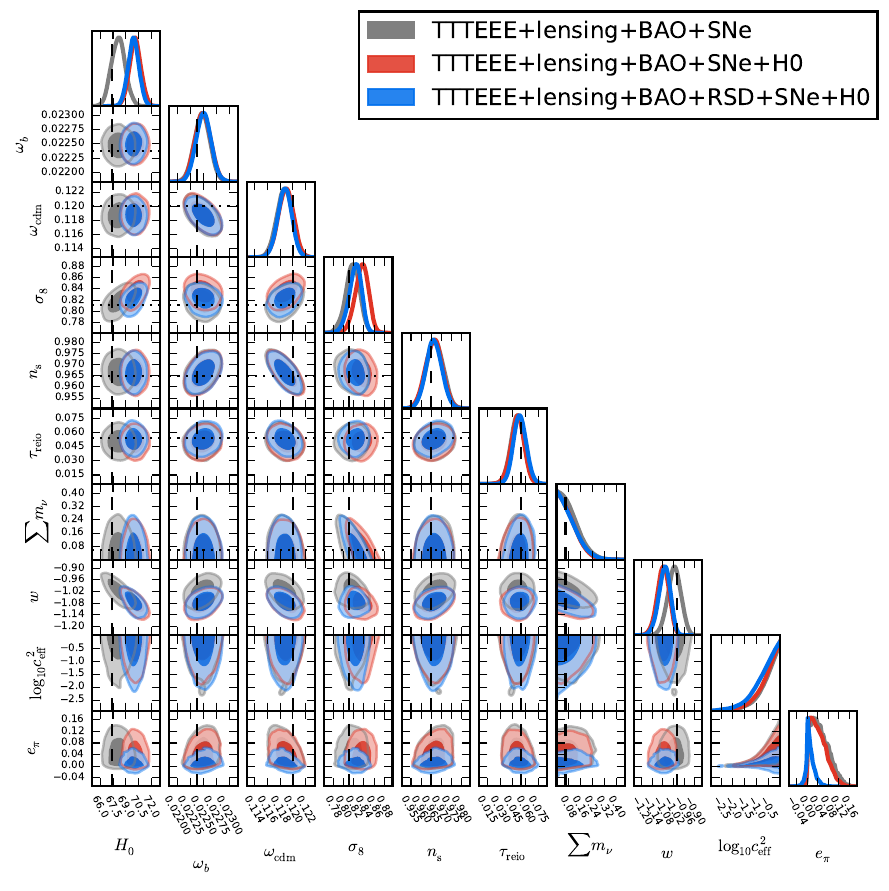}
\caption{The 1-D and 2-D posteriors for the cosmological parameters in the ADE model inferred from several data sets; we show $68\%$ and $95\%$ confidence contours. The points where dashed vertical lines and dotted horizontal lines meet denote $\Lambda$CDM baseline result reported by the Planck collaboration. This analysis uses the flat prior bounds in Table~\ref{Table:prior-MGmodels}. Note that here we plot the Hubble constant $H_0$ and the strength of matter clustering $\sigma_8$ which are derived parameters in our analysis.}
\label{Fig:constraints-MI}
\end{figure*}

We present our results in Fig.~\ref{Fig:constraints-MI} and Table~\ref{Table:constraints-MI}. Note that, since we performed several probe combinations, in order to avoid overcrowding in Fig.~\ref{Fig:constraints-MI} we only display three relevant cases. Our constraints for the parameters in common with the standard cosmological model $\Lambda$CDM are in good agreement with the results found by the Planck Collaboration when the prior on $H_0$ is not used. Including a prior in the Hubble constant drives our result for $H_0$ towards high values, while also favouring values of the equation of state in the phantom regime $w<-1$. Whereas RSD play an important part in reducing the uncertainty of the anisotropic stress parameter $\epi$, the DE sound speed remains unconstrained  whatever probe combination we utilise.  

\begin{table}[h]
\centering
\resizebox{\textwidth}{!}{%
\begin{tabular}{c c c c c c c c}
\hline
Parameter & \texttt{P}+$H_0$  & \texttt{P}+BAO & \texttt{P}+SNe & \texttt{P}+BAO+$H_0$ & \texttt{P}+BAO+SNe & $\left\lbrace\dots\right\rbrace$+$H_0$ & $\left\lbrace\dots\right\rbrace$+RSD\\
\hline
$\omega_{\rm{b}}$ & $ 0.02240^{-0.00016}_{+0.00017} $ & $0.02248\pm 0.00016$ & $0.02240\pm 0.00017 $ & $0.02239\pm 0.00015 $ & $0.02248\pm 0.00015$ & $0.02249\pm 0.00015$ & $0.02250\pm 0.00014 $ \\
$\omega_{\rm{cdm}}$ & $0.1195\pm 0.0014$ & $0.1186\pm 0.0013 $ & $0.1195\pm 0.0014$ & $0.1198\pm 0.0012 $ & $0.1187\pm 0.0012$ & $0.1189\pm 0.0012$ & $0.1187\pm 0.0011 $\\
$H_0\,(\frac{\rm{km}}{\rm{s \cdot Mpc}})$ & $73.00\pm 1.06$ & $68.03^{-1.63}_{+1.35}$ & $67.74^{-1.28}_{+1.24}$ & $71.70^{-0.90}_{+0.92}$ & $68.12^{-0.81}_{+0.78}$ & $70.09^{-0.67}_{+0.68}$ & $69.89^{-0.64}_{+0.63}$ \\
$ \ln 10^{10}A_{\rm{s}}$ & $3.034^{-0.015}_{+0.016}$ & $3.033^{-0.015}_{+0.016}$ & $3.033\pm 0.016$ & $3.034\pm 0.015 $ & $3.032\pm 0.016 $ & $3.032\pm 0.016$ & $3.039\pm 0.015$ \\
$n_{\rm{s}}$ & $0.9642\pm 0.0048$ & $0.9670\pm 0.0044$ & $0.9643\pm 0.0050 $ & $0.9637\pm 0.0042$ & $0.9668\pm 0.0042$ & $0.9664^{-0.0043}_{+0.0042}$ & $0.9664\pm 0.0039$ \\
$\tau_{\rm{reio}} $ & $0.0502^{-0.0076}_{+0.0077}$ & $0.0507^{-0.0078}_{+0.0079}$ & $0.0503^{-0.0075}_{+0.0074}$ & $0.0501^{-0.0074}_{+0.0073}$ & $0.0504^{-0.0078}_{0.0079} $ & $0.0503^{-0.0073}_{+0.0080}$ & $0.0531^{-0.0074}_{+0.0075}$ \\
$\sum m_\nu\,(\rm{eV}) $ & $ <0.212$ & $<0.117$ & $<0.221$ & $ 0.162^{-0.123}_{+0.070} $ & $<0.111$ & $<0.102$ & $<0.105 $ \\
$w$ & $-1.22^{-0.05}_{+0.08}$ & $-1.01^{-0.06}_{+0.08}$ & $-1.04^{-0.04}_{+0.05}$ & $-1.18^{-0.05}_{+0.07} $ & $ -1.02\pm 0.04$ & $ -1.08^{-0.03}_{+0.04} $ & $-1.07^{-0.03}_{+0.04} $ \\ 
$\log_{10} \ceff^2$ & $ >-0.6 $ & $ >-0.7 $ & $ >-0.7 $ & $ >-0.6 $ & $ >-0.6 $ & $ >-0.6$ & $>-0.8$ \\
$\epi$ & $ 0.042^{-0.040}_{+0.017}$ & $0.049^{-0.047}_{+0.018}$ & $0.049^{-0.046}_{+0.018}$ & $ 0.041^{-0.040}_{+0.016} $ & $ 0.050^{-0.049}_{+0.016}$ & $0.044^{-0.041}_{+0.017}$ & $0.012^{-0.015}_{+0.011}$ \\
$\sigma_8 $ & $ 0.850^{-0.016}_{+0.023} $ & $ 0.816\pm 0.017 $ & $0.806^{-0.017}_{+0.027} $ & $0.842^{-0.015}_{+0.017}$ & $0.818^{-0.013}_{+0.014}$  & $0.836^{-0.012}_{+0.014}$ & $0.823^{-0.011}_{+0.014}$ \\
\hline
\end{tabular}}
\caption{Mean values and $68\%$ confidence limits on cosmological parameters for the ADE model. Here $\left\lbrace\dots\right\rbrace$ stands for the inclusion of data from column on the left and \texttt{P} stands for the inclusion of Planck data (i.e., temperature and polarisation anisotropies of the CMB as well as CMB lensing).}
\label{Table:constraints-MI}
\end{table}

\section{Forecast}
\label{sec:forecast}

\subsection{Methodology}
\label{sec:methodology}

Analyses of ongoing and forthcoming galaxy surveys will be of paramount importance for cosmology. Fluctuations in the number counts have different systematics as compared to CMB fluctuations, hence their measurements can break degeneracies and certainly improve cosmological constraints~\cite{Eisenstein:1998hr}. In the usual approach galaxy number counts are compared to the predicted matter power spectrum of matter density fluctuations $P(k,z)$. This quantity has however a few disadvantages. 

Firstly, it is not directly observable and assumptions are made when dealing with data~\cite{Bonvin:2011bg}. Since galaxy surveys measure both redshifts and angles one must assume a distance--redshift relation, which depends on cosmological parameters such as $\Omega_m$, to have the data points in physical space (as opposed to redshift space) where it is possible to compute the power spectrum. Secondly, it is not trivial to include lensing effects in the standard matter power spectrum approach $P(k,z)$ because lensing inherently mixes different scales.\footnote{See, however, Ref.~\cite{Castorina:2021xzs} where the authors showed that the impact of lensing magnification on the multipoles of the observed power spectrum can reach $5-10\%$ of the monopole and the quadrupole at large scales for sufficiently realistic high redshift surveys. Since the main difference between the approaches is the assumed fiducial model required to convert angles and redshifts into distances (the power spectrum in harmonic space does not make such an assumption), we expect that as long as the assumed fiducial cosmology closely matches the true underlying one, the results computed in the two approaches should be quite similar.} The alternative approach using the power spectrum in harmonic space $C_\ell(z,z')$ might avoid these drawbacks, as this approach makes no model assumptions in dealing with data and the power spectrum in harmonic space is an observable~\cite{DiDio:2013bqa}. Moreover, relevant relativistic effects such as lensing convergence and RSD are easily included~\cite{Bonvin:2011bg,DiDio:2013bqa}. In addition, an analysis carried out with $C_\ell(z,z')$ is frame independent~\cite{Francfort:2019ynz}.

In this paper we use the power spectrum in harmonic space $C_\ell(z,z')$ to estimate the bias in  cosmological parameters due to neglecting lensing convergence when performing analyses of ongoing and upcoming galaxy surveys. We follow the approach in Ref.~\cite{Cardona:2016qxn}  and compute the power spectrum in harmonic space with our modified code \texttt{CLASS}. Overall, the procedure is as follows:
\begin{itemize}
\item For a given fiducial model, we compute the ``observed'' $C_\ell^{\rm obs}$ which include matter perturbations, RSD, and lensing convergence.
\item We carry out Markov Chain Monte Carlo (MCMC) analyses using the ``theory'' $C_\ell^{\rm{th}}$ in two cases: i) consistently including lensing convergence when modelling number counts fluctuations and; ii) neglecting lensing convergence.
\end{itemize}
The code \texttt{CLASS} requires survey specifications (e.g., number of galaxies per redshift and per steradian, galaxy density, magnification bias, covered sky fraction, galaxy bias) to compute the power spectrum in harmonic space. In this work we will utilise a survey configuration which is consistent with the Euclid photometric catalogue. These survey specifications were given in Appendix A of Ref.~\cite{Cardona:2016qxn} and our implementation is exactly the same.

We specify the ADE cosmological model by the following parameters:  $\omega_{\rm{b}}$, $\omega_{\rm{cdm}}$, $n_{\rm{s}}$, $\ln 10^{10}A_{\rm{s}}$, $H_0$, $\sum m_\nu$, $\tau$, $w$, $\log_{10} \ceff^2$, $\epi$, and $b_0$. The latter is the amplitude of the scale-independent galaxy bias prescription that we assume, namely,  
\begin{equation}
b(z) = b_0 \sqrt{1+z}.
\label{Eq:bias}
\end{equation}
As for the fiducial ADE model, we assume our baseline constraint having parameter values given by the last column in Table~\ref{Table:constraints-MI} (i.e., $\left\lbrace\dots\right\rbrace$+RSD). The fiducial galaxy bias amplitude is set to $b_0=1$.   

In order to carry out the statistical analysis for our forecast we take into consideration a few additional points that we now briefly discuss. First, since galaxy number counts are discrete tracers of the underlying dark matter distribution, it is necessary to take into account Poisson shot noise in our analysis. Second, an additional source of error is our relative ignorance about the non--linear behaviour of number counts fluctuations. We consider this uncertainty by adding a non--linear error term computed as
\begin{equation}
E_\ell^{ij} = | C_\ell^{ij,\rm{HALOFIT\,ON}} - C_\ell^{ij,\rm{HALOFIT\,OFF}} |.
\end{equation}
Therefore, we model the angular power spectrum of number counts fluctuations as
\begin{equation}
C_\ell^{{\rm A},\,ij} = C_\ell^{{\rm A},\,ij} + E_\ell^{ij} + \mathcal{N}^{-1} \delta_{ij},
\label{Eq:angular-power-spectrum}
\end{equation}
where $\mathrm{A=obs,th}$, $C_\ell^{{\rm A},\,ij}$ on the right-hand side is computed by \texttt{CLASS}, $i,j=1,...,N_{\rm bin}$ are redshift bin indices, $E_\ell^{ij}$ denotes the non--linear error term, $\mathcal{N}$ is the number of galaxies per steradian. Here `obs' and `th' respectively stand for `observed' and `theory', whereas $N_{\rm bin}$ is the number of redshift bins. Third, note that the computation of number counts spectra requires a considerable amount of computational resources rapidly increasing with the number of bins if computed accurately. In order to speed up our analysis and make it doable, we proceed as follows. While we compute $C_\ell^{\mathrm{obs},ij}$ and $E_\ell^{ij}$ accurately only once, we use less precise\footnote{We find precision parameters for the code such that at the fiducial model $\Delta \chi^2 \lesssim 1$.} $C_\ell^{\mathrm{th},ij}$ which are computed $\sim 10^5$ times. \texttt{CLASS} precision parameters for computation of $C_\ell^{\mathrm{obs},ij}$ and $C_\ell^{\mathrm{th},ij}$ are released with our modified version of the code. We adopted the Halofit model \cite{takahashi2012revising} to describe the non-linear matter power spectrum.
Exploring Large-Scale Structure (LSS) inherently requires methodologies extending beyond linear order approximations. See for example \cite{di2014galaxy,yoo2014beyond} for insights into second-order galaxy number counts. Determining the non-linear power spectrum in the context of GR remains an unresolved challenge, particularly when Poisson equations undergo modifications, as seen in MG theories, like the models we are considering. One approach to address this issue involves computing the progression of matter perturbations through an N-body simulation \cite{takahashi2012revising}. Nevertheless, this method proves to be labor-intensive and entails significant computational costs, hence we leave this for follow-up work.

In our MCMC analysis we follow\footnote{Note that the parameter dependence of the non--linear error term is neglected, that is, $E_\ell^{ij}$ in Eq.~\eqref{Eq:angular-power-spectrum} is  only computed for the fiducial model.} Ref.~\cite{Cardona:2016qxn}. We use wide flat priors unless we specify it differently and implement a Gaussian likelihood which allows us to compute a $\chi^2$ relative to the fiducial model given by
\begin{equation}
\Delta \chi^2 = \sum_{\ell=2}^{\ell_{\max}} (2\ell+1) f_{\rm sky} \left( \ln \frac{d_\ell^{\rm th}}{d_\ell^{\rm obs}} + \frac{d_\ell^{\rm mix}}{d_\ell^{\rm th}} - N_{\rm bin}\right),
\label{eq:chi2}
\end{equation}
where $f_{\rm sky} $ is the covered sky fraction, $d_\ell^{\mathrm{A}} \equiv \det ( C_\ell^{\mathrm{A},ij} )$ and $d_\ell^{\rm mix}$ is computed like $d_\ell^{\rm th}$ but substituting in each term of the determinant  one factor by $C_\ell^{\mathrm{obs},ij}$.  The total  $d_\ell^{\rm mix}$ is obtained by adding all different possibilities for the insertion of $C_\ell^{\mathrm{obs},ij}$.\footnote{More details and an example can be found in Ref.~\cite{Audren:2012vy}.} To be conservative and keep non--linear effects under control, we choose $\ell_{\rm max}=400$ in the analysis. Note that we limit our analysis to $\ell \leq \ell_{\rm max}$ so that scales $\lambda_{\rm min} \leq 2\pi/k_{\rm max}$ and orthogonal to the line-of-sight, namely where non-linearities start to play a part, are not considered. In flat space this smallest scale condition is equivalent to $[2\pi/\ell_{\rm max}]D_{\rm A}(\bar{z})= a(\bar{z})[2\pi/k_{\rm max}]$, yielding $\ell_{\rm max} = r(\bar{z})k_{\rm max}$ with $D_{\rm A}$ and $r$ the angular diameter distance and comoving radius, respectively. Moreover, in order to take into account possible effects of different galaxy selection functions, we carry out the forecast for two cases (i.e., top-hat and Gaussian) each having the number of redshift bins $N_{\rm bin}=5,10$.

\subsection{Results and discussion}
\label{sec:model-I}

We perform the forecast analysis also including information from our cosmological constraints in Section~\ref{sec:constraints}. From the chains for our baseline constraint (last column in Table~\ref{Table:constraints-MI}) we compute the covariance matrix $\mathbf{C}$ for the parameters  $\vec{\mathbf{x}}=(\omega_{\rm{b}},\, \omega_{\rm{cdm}},\, n_{\rm{s}},\, \ln 10^{10}A_{\rm{s}},\, H_0,\, \tau,\, w,\, \epi)$ which are well constrained by available data sets. We then perform forecasts assuming  a Gaussian distribution for the prior $\vec{\mathbf{x}}$ and wide, flat distributions for the remaining cosmological parameters (i.e., $\sum m_\nu$, $b_0$, $\log_{10} \ceff^2$). The $\chi^2$ relative to the fiducial model including the Gaussian prior is then the $\Delta \chi^2$ in Eq. \eqref{eq:chi2} plus
\begin{equation}
\Delta \chi^2_{\mathrm{prior}} = \sum_{i,j} (x_i - x^{\mathrm{fid}}_i) C^{-1}_{ij} (x_j - x^{\mathrm{fid}}_j),
\label{Eq:chi2-prior}
\end{equation}
where $\vec{x}^{\mathrm{fid}}$ denotes parameters of the fiducial model and $\mathbf{C^{-1}}$ is the inverse of the covariance matrix $\mathbf{C}$. We determine the bias of the cosmological parameters due to neglecting lensing convergence by fitting the fiducial $C^{\rm obs}_\ell$ with $C^{\rm th}_\ell$ where lensing convergence is i) consistently included and ii) wrongly neglected.

\begin{figure*}[h]
\centering
\includegraphics[width=\textwidth]{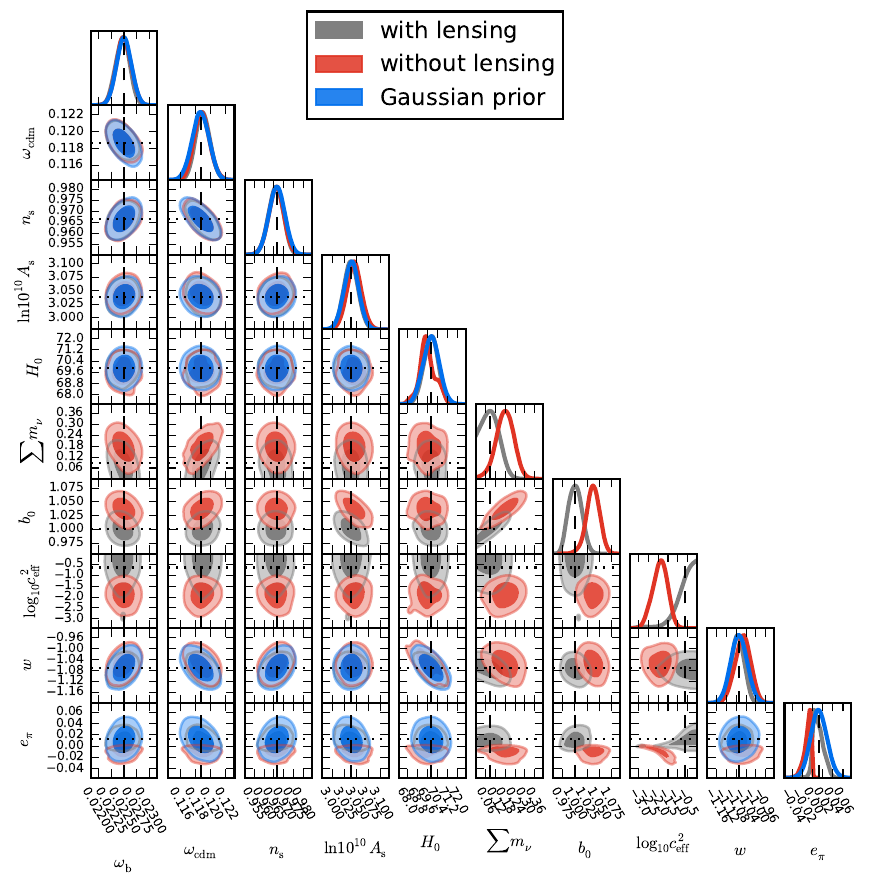}
\caption{Results for survey configuration of 5 Gaussian redshift bins. The 1-D and 2-D posteriors for the cosmological parameters in the ADE model inferred from a consistent analysis including lensing convergence (gray) and an analysis neglecting lensing convergence (red). We show $68\%$ and $95\%$ confidence contours. The points where dashed vertical lines and dotted horizontal lines meet denote the fiducial cosmology, namely, our baseline constraint. This analysis uses information from the constraints presented in Sec.~\ref{sec:constraints} for the parameters $\omega_{\rm b}$, $\omega_{\rm cdm}$, $n_{\rm s}$, $\ln 10^{10}A_{\rm s}$, $H_0$, $w$, and $\epi$ as indicated by the Gaussian prior in blue.}
\label{Fig:g5}
\end{figure*}

\begin{figure*}[h]
\centering
\includegraphics[width=\textwidth]{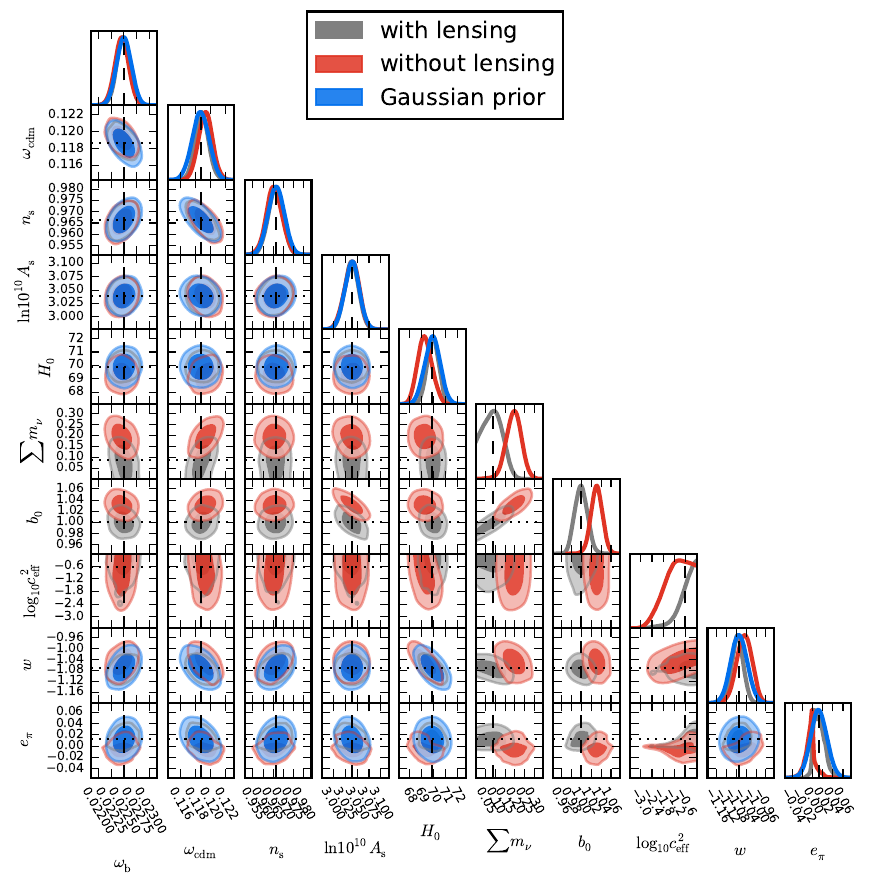}
\caption{Results for survey configuration of 10 Gaussian redshift bins. The 1-D and 2-D posteriors for the cosmological parameters in the ADE model inferred from a consistent analysis including lensing convergence (gray) and an analysis neglecting lensing convergence (red). We show $68\%$ and $95\%$ confidence contours. The points where dashed vertical lines and dotted horizontal lines meet denote the fiducial cosmology, namely, our baseline constraint. This analysis uses information from the constraints presented in Sec.~\ref{sec:constraints} for the parameters $\omega_{\rm b}$, $\omega_{\rm cdm}$, $n_{\rm s}$, $\ln 10^{10}A_{\rm s}$, $H_0$, $w$, and $\epi$ as indicated by the Gaussian prior in blue.}
\label{Fig:g10}
\end{figure*}

\begin{figure*}[h]
\centering
\includegraphics[width=\textwidth]{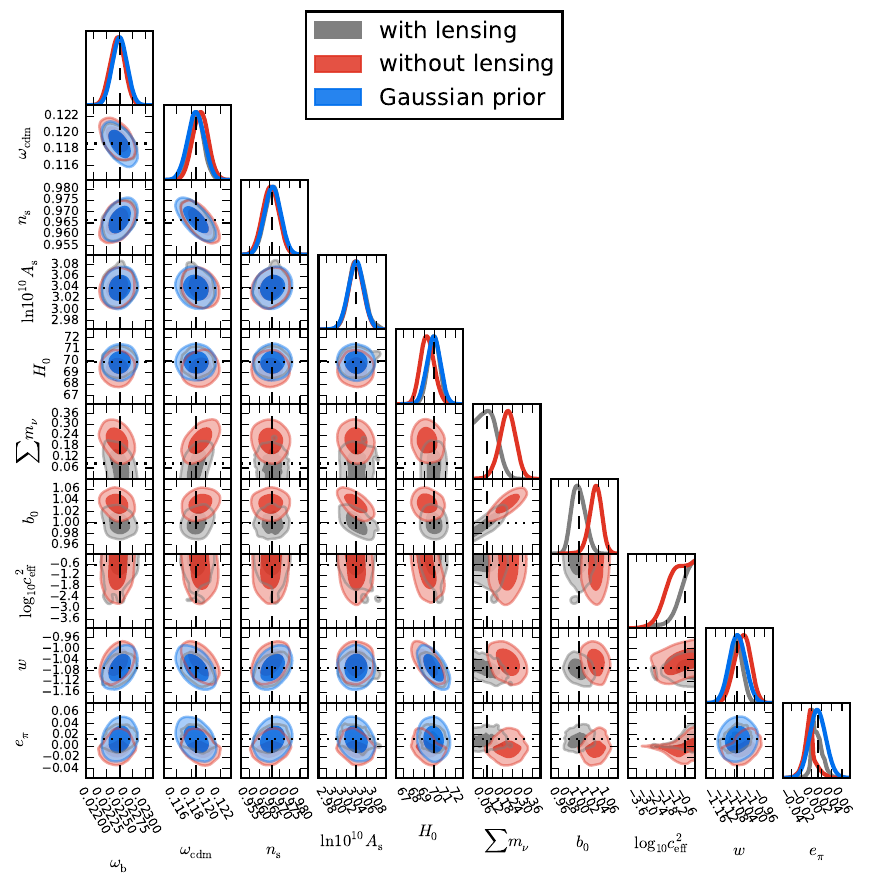}
\caption{Results for survey configuration of 5 top-hat redshift bins. The 1-D and 2-D posteriors for the cosmological parameters in the ADE model inferred from a consistent analysis including lensing convergence (gray) and an analysis neglecting lensing convergence (red). We show $68\%$ and $95\%$ confidence contours. The points where dashed vertical lines and dotted horizontal lines meet denote the fiducial cosmology, namely, our baseline constraint. This analysis uses information from the constraints presented in Sec.~\ref{sec:constraints} for the parameters $\omega_{\rm b}$, $\omega_{\rm cdm}$, $n_{\rm s}$, $\ln 10^{10}A_{\rm s}$, $H_0$, $w$, and $\epi$ as indicated by the Gaussian prior in blue.}
\label{Fig:t5}
\end{figure*}

\begin{figure*}[h]
\centering
\includegraphics[width=\textwidth]{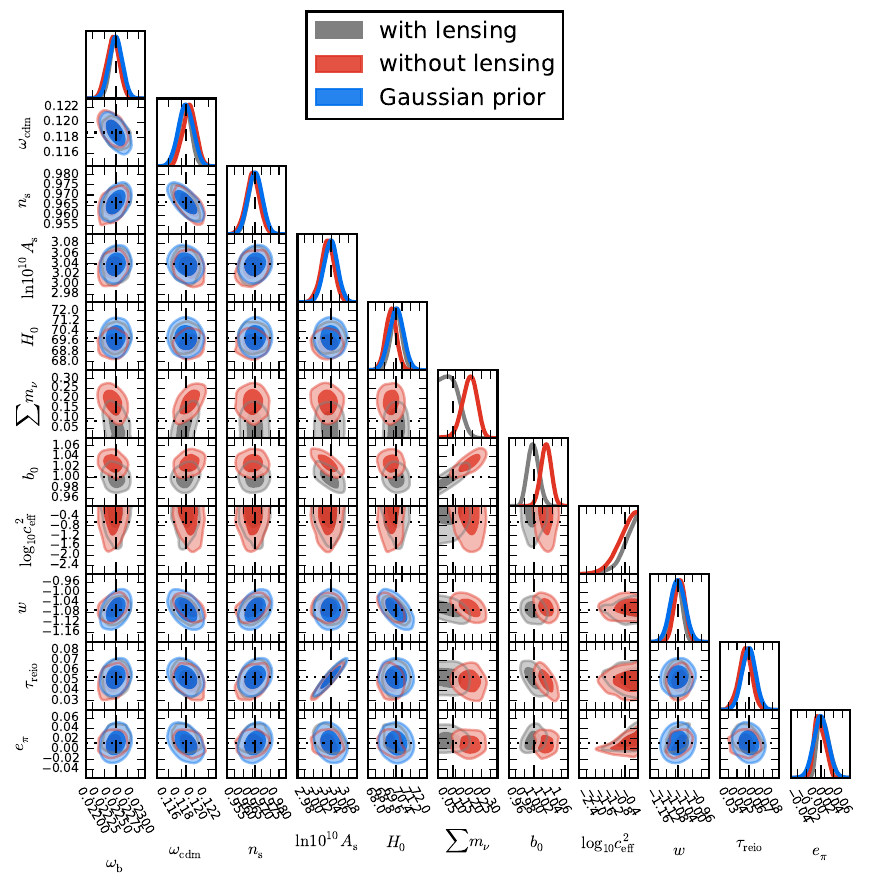}
\caption{Results for survey configuration of 10 top-hat redshift bins. The 1-D and 2-D posteriors for the cosmological parameters in the ADE model inferred from a consistent analysis including lensing convergence (gray) and an analysis neglecting lensing convergence (red). We show $68\%$ and $95\%$ confidence contours. The points where dashed vertical lines and dotted horizontal lines meet denote the fiducial cosmology, namely, our baseline constraint. This analysis uses information from the constraints presented in Sec.~\ref{sec:constraints} for the parameters $\omega_{\rm b}$, $\omega_{\rm cdm}$, $n_{\rm s}$, $\ln 10^{10}A_{\rm s}$, $H_0$, $w$, and $\epi$ as indicated by the Gaussian prior in blue.}
\label{Fig:t10}
\end{figure*}

Our results are shown in Fig.~\ref{Fig:g5} and Table~\ref{Table:g5} for 5  Gaussian redshift bins; Fig.~\ref{Fig:g10} and Table~\ref{Table:g10} for 10 Gaussian redshift bins; Fig.~\ref{Fig:t5} and Table~\ref{Table:t5} for 5 top-hat redshift bins; Fig.~\ref{Fig:t10} and Table~\ref{Table:t10} for 10 top-hat redshift bins. Figs.~\ref{Fig:g5}-\ref{Fig:t10} show  $68\%$ and $95\%$ confidence contours for a model consistently including lensing convergence (gray) and for a model neglecting lensing convergence (red) when modelling number counts fluctuations; the vertical, dashed lines and the horizontal, dotted lines indicate parameter values in our fiducial model.

\begin{table}[h]
\centering
\begin{tabular}{c c c c c c}
\hline
\multicolumn{6}{c}{i) Consistently including lensing: $\Delta \chi^2 = 1$} \\
\hline
Parameter & Mean & Best fit & $\sigma$ &\hspace{-0.52cm} shift: Mean & Best fit \\
\hline
$\omega_{\rm{b}}$ & $0.02247$ & $0.02256 $ & $0.00013 $ &  \quad$-0.2\sigma$ & $ 0.4\sigma$ \\
$\omega_{\rm{cdm}}$ & $0.1189 $ & $0.1183 $ & \quad$0.0010 $ &  \quad$0.2\sigma$ & $-0.5\sigma$ \\
$n_{\rm{s}}$      & $0.9660 $ & $0.9676 $ & $0.0037 $ &  \quad$-0.1\sigma$ & $ 0.3\sigma$ \\
$\ln10^{10}A_{\rm{s}}$ & $3.038 $ & $3.043$ & $0.015 $ &  \quad$<|0.1\sigma|$ & $ 0.3\sigma$ \\
$H_0\left(\frac{\text{km}}{\text{s}\cdot\text{Mpc}}\right)$ & $69.87$ & $70.12$ & $0.56$ &  \quad$<|0.1\sigma|$ & $ 0.4\sigma$ \\
$\sum m_{\nu}$\,(eV)  & $0.09$ & $0.04$ & $0.05$ & \quad $ 0.1\sigma$ & $ -0.8\sigma$ \\
$b_0$ & $1.001$ & $0.990$ & $0.013$ & \quad$0.1\sigma$ & $-0.7\sigma$ \\
$\log_{10} \ceff^2$ & $-0.647$ & $-0.671$ & $0.557$ & \quad$<|0.1\sigma|$ & $-0.1\sigma$ \\
$w$ & $-1.08$ & $-1.07$ & $0.03$ & \quad$-0.2\sigma$ & $0.1\sigma$ \\
$\epi$ & $0.011$ & $0.012$ & $0.010$ & \quad$-0.1\sigma$ & $<|0.1\sigma|$ \\	
\end{tabular}
\begin{tabular}{c c c c c c}
\hline
\multicolumn{6}{c}{ii) Neglecting lensing: $\Delta \chi^2 = 1911$} \\
\hline
Parameter & Mean & Best fit & $\sigma$ & \hspace{-0.52cm} shift: Mean & Best fit \\
\hline
$\omega_{\rm{b}}$ & $0.02249$ & $0.02236 $ & $0.00014 $ &  \quad$<|0.1\sigma|$ & $-\sigma$ \\
$\omega_{\rm{cdm}}$ & $0.1188$ & $0.1187$ & $0.0010$ &  \quad$0.1\sigma$ & $<0.1\sigma$ \\
$n_{\rm{s}}$      & $0.9662$ & $0.9649$ & $0.0038$ &  \quad$-0.1\sigma$ & $-0.4\sigma$ \\
$\ln10^{10}A_{\rm{s}}$ & $ 3.045 $ & $3.043 $ & $ 0.015 $ &  \quad$0.4\sigma$ & $0.3\sigma$ \\
$H_0\left(\frac{\text{km}}{\text{s}\cdot\text{Mpc}}\right)$      & $69.61$ & $69.11$ & $0.65$ &  \quad$-0.4\sigma$ & $-1.2\sigma$ \\
$\sum m_{\nu}$\,(eV)  & $0.18$ & $0.20$ & $0.05$ &  \quad$1.7\sigma$ & $2.1\sigma$ \\
$b_0$ & $1.037$ & $1.040$ & $0.014$ & \quad$2.7\sigma$ & $3.\sigma$\\
$\log_{10} \ceff^2$ & $-1.907$ & $-1.866$ & $0.386$ & \quad$-3.3\sigma$ & $-3.2\sigma$ \\
$w$ & $-1.05$ & $-1.04$ & $0.03$ & \quad$0.5\sigma$ & $0.9\sigma$ \\
$\epi$ & $-0.012$ & $-0.007$ & $0.007$ & \quad$-3\sigma$ & $-2.5\sigma$ \\	
\hline
\end{tabular}
\caption{Results for survey configuration of 5 Gaussian redshift bins. MCMC results for ADE model when considering information from the constraints presented in Sec.~\ref{sec:constraints} for the parameters $\omega_{\rm b}$, $\omega_{\rm cdm}$, $n_{\rm s}$, $\ln 10^{10}A_{\rm s}$, $H_0$, $w$, and $\epi$. We show the mean, the best fit, the standard deviation, and the shift of both mean and best fit with respect to the fiducial values in units of the standard deviation. Because the theoretical spectra are computed less accurately than the observed spectra,  deviations smaller than $0.2\sigma$ are not significant.}
\label{Table:g5}
\end{table}

\begin{table}[h]
\centering
\begin{tabular}{c c c c c c}
\hline
\multicolumn{6}{c}{i) Consistently including lensing: $\Delta \chi^2 = 1$} \\
\hline
Parameter & Mean & Best fit & $\sigma$ &\hspace{-0.52cm} shift: Mean & Best fit \\
\hline
$\omega_{\rm{b}}$ & $0.02249$ & $0.02247 $ & $0.00013 $ &  \quad$-0.1\sigma$ & $-0.3\sigma$ \\
$\omega_{\rm{cdm}}$ & $0.1188 $ & $0.1183 $ & \quad$0.0009 $ &  \quad$0.1\sigma$ & $-0.4\sigma$ \\
$n_{\rm{s}}$      & $0.9662 $ & $0.9683 $ & $0.0036 $ &  \quad$-0.1\sigma$ & $ 0.5\sigma$ \\
$\ln10^{10}A_{\rm{s}}$ & $3.038 $ & $3.041$ & $0.015 $ &  \quad$<|0.1\sigma|$ & $ 0.1\sigma$ \\
$H_0\left(\frac{\text{km}}{\text{s}\cdot\text{Mpc}}\right)$ & $69.93$ & $69.77$ & $0.49$ &  \quad$0.1\sigma$ & $-0.2\sigma$ \\
$\sum m_{\nu}$\,(eV)  & $0.09$ & $0.10$ & $0.05$ & \quad $<0.1\sigma$ & $ 0.3\sigma$ \\
$b_0$ & $1.000$ & $1.003$ & $0.013$ & \quad$<|0.1\sigma|$ & $0.2\sigma$ \\
$\log_{10} \ceff^2$ & $-0.580$ & $-0.690$ & $0.483$ & \quad$0.1\sigma$ & $-0.1\sigma$ \\
$w$ & $-1.08$ & $-1.07$ & $0.02$ & \quad$-0.2\sigma$ & $0.2\sigma$ \\
$\epi$ & $0.013$ & $0.012$ & $0.011$ & \quad$0.1\sigma$ & $<0.1\sigma$ \\	
\end{tabular}
\begin{tabular}{c c c c c c}
\hline
\multicolumn{6}{c}{ii) Neglecting lensing: $\Delta \chi^2 = 2564$} \\
\hline
Parameter & Mean & Best fit & $\sigma$ & \hspace{-0.52cm} shift: Mean & Best fit \\
\hline
$\omega_{\rm{b}}$ & $0.02246$ & $0.02239 $ & $0.00013 $ &  \quad$-0.3\sigma$ & $-0.8\sigma$ \\
$\omega_{\rm{cdm}}$ & $0.1192$ & $0.1196$ & $0.0009$ &  \quad$0.6\sigma$ & $\sigma$ \\
$n_{\rm{s}}$      & $0.9655$ & $0.9661$ & $0.0039$ &  \quad$-0.2\sigma$ & $-0.1\sigma$ \\
$\ln10^{10}A_{\rm{s}}$ & $ 3.039 $ & $3.049 $ & $ 0.016 $ &  \quad$<|0.1\sigma|$ & $0.7\sigma$ \\
$H_0\left(\frac{\text{km}}{\text{s}\cdot\text{Mpc}}\right)$ & $69.26$ & $69.20$ & $0.60$ &  \quad$-\sigma$ & $-1.1\sigma$ \\
$\sum m_{\nu}$\,(eV)  & $0.19$ & $0.20$ & $0.04$ &  \quad$2.7\sigma$ & $2.8\sigma$ \\
$b_0$ & $1.031$ & $1.028$ & $0.011$ & \quad$2.7\sigma$ & $2.4\sigma$\\
$\log_{10} \ceff^2$ & $-1.061$ & $-1.605$ & $0.660$ & \quad$-0.6\sigma$ & $-1.5\sigma$ \\
$w$ & $-1.05$ & $-1.05$ & $0.03$ & \quad$0.7\sigma$ & $0.5\sigma$ \\
$\epi$ & $-0.005$ & $-0.007$ & $0.011$ & \quad$-1.6\sigma$ & $-1.8\sigma$ \\	
\hline
\end{tabular}
\caption{Results for survey configuration of 10 Gaussian redshift bins. MCMC results for ADE model when considering information from the constraints presented in Sec.~\ref{sec:constraints} for the parameters $\omega_{\rm b}$, $\omega_{\rm cdm}$, $n_{\rm s}$, $\ln 10^{10}A_{\rm s}$, $H_0$, $w$, and $\epi$. We show the mean, the best fit, the standard deviation, and the shift of both mean and best fit with respect to the fiducial values in units of the standard deviation. Because the theoretical spectra are computed less accurately than the observed spectra,  deviations smaller than $0.2\sigma$ are not significant.}
\label{Table:g10}
\end{table}

\begin{table}[h]
\centering
\begin{tabular}{c c c c c c}
\hline
\multicolumn{6}{c}{i) Consistently including lensing: $\Delta \chi^2 = 0$} \\
\hline
Parameter & Mean & Best fit & $\sigma$ &\hspace{-0.52cm} shift: Mean & Best fit \\
\hline
$\omega_{\rm{b}}$ & $0.02249$ & $0.02250 $ & $0.00014 $ &  \quad$-0.1\sigma$ & $ <0.1\sigma$ \\
$\omega_{\rm{cdm}}$ & $0.1188 $ & $0.1187 $ & \quad$0.0009 $ &  \quad$0.1\sigma$ & $<0.1\sigma$ \\
$n_{\rm{s}}$      & $0.9664 $ & $0.9664 $ & $0.0037 $ &  \quad$<|0.1\sigma|$ & $ <0.1\sigma$ \\
$\ln10^{10}A_{\rm{s}}$ & $3.041 $ & $3.039$ & $0.016 $ &  \quad$0.1\sigma$ & $ <0.1\sigma$ \\
$H_0\left(\frac{\text{km}}{\text{s}\cdot\text{Mpc}}\right)$ & $69.91$ & $69.89$ & $0.56$ &  \quad$<0.1\sigma$ & $ <0.1\sigma$ \\
$\sum m_{\nu}$\,(eV)  & $0.09$ & $0.09$ & $0.05$ & \quad $ 0.1\sigma$ & $ <0.1\sigma$ \\
$b_0$ & $1.000$ & $1.000$ & $0.013$ & \quad$<|0.1\sigma|$ & $<0.1\sigma$ \\
$\log_{10} \ceff^2$ & $-0.696$ & $-0.640$ & $0.569$ & \quad$-0.1\sigma$ & $<0.1\sigma$ \\
$w$ & $-1.08$ & $-1.07$ & $0.03$ & \quad$-0.2\sigma$ & $<0.1\sigma$ \\
$\epi$ & $0.011$ & $0.012$ & $0.011$ & \quad$-0.1\sigma$ & $<0.1\sigma$ \\	
\end{tabular}
\begin{tabular}{c c c c c c}
\hline
\multicolumn{6}{c}{ii) Neglecting lensing: $\Delta \chi^2 = 1734$} \\
\hline
Parameter & Mean & Best fit & $\sigma$ & \hspace{-0.52cm} shift: Mean & Best fit \\
\hline
$\omega_{\rm{b}}$ & $0.02246$ & $0.02243 $ & $0.00014 $ &  \quad$-0.3\sigma$ & $-0.5\sigma$ \\
$\omega_{\rm{cdm}}$ & $0.1192$ & $0.1194$ & $0.0011$ &  \quad$0.5\sigma$ & $0.6\sigma$ \\
$n_{\rm{s}}$      & $0.9655$ & $0.9658$ & $0.0039$ &  \quad$-0.2\sigma$ & $-0.1\sigma$ \\
$\ln10^{10}A_{\rm{s}}$ & $ 3.039 $ & $3.041 $ & $ 0.015 $ &  \quad$<0.1\sigma$ & $0.2\sigma$ \\
$H_0\left(\frac{\text{km}}{\text{s}\cdot\text{Mpc}}\right)$ & $69.30$ & $69.14$ & $0.66$ &  \quad$-0.9\sigma$ & $-1.1\sigma$ \\
$\sum m_{\nu}$\,(eV)  & $0.21$ & $0.21$ & $0.05$ &  \quad$2.5\sigma$ & $2.5\sigma$ \\
$b_0$ & $1.034$ & $1.035$ & $0.012$ & \quad$2.8\sigma$ & $2.8\sigma$\\
$\log_{10} \ceff^2$ & $-1.062$ & $-1.722$ & $0.687$ & \quad$-0.6\sigma$ & $-1.6\sigma$ \\
$w$ & $-1.05$ & $-1.05$ & $0.03$ & \quad$0.6\sigma$ & $0.5\sigma$ \\
$\epi$ & $-0.004$ & $-0.005$ & $0.012$ & \quad$-1.4\sigma$ & $-1.4\sigma$ \\	
\hline
\end{tabular}
\caption{Results for survey configuration of 5 top-hat redshift bins. MCMC results for ADE model when considering information from the constraints presented in Sec.~\ref{sec:constraints} for the parameters $\omega_{\rm b}$, $\omega_{\rm cdm}$, $n_{\rm s}$, $\ln 10^{10}A_{\rm s}$, $H_0$, $w$, and $\epi$. We show the mean, the best fit, the standard deviation, and the shift of both mean and best fit with respect to the fiducial values in units of the standard deviation. Because the theoretical spectra are computed less accurately than the observed spectra,  deviations smaller than $0.2\sigma$ are not significant.}
\label{Table:t5}
\end{table}

\begin{table}[h]
\centering
\begin{tabular}{c c c c c c}
\hline
\multicolumn{6}{c}{i) Consistently including lensing: $\Delta \chi^2 = 1$} \\
\hline
Parameter & Mean & Best fit & $\sigma$ &\hspace{-0.52cm} shift: Mean & Best fit \\
\hline
$\omega_{\rm{b}}$ & $0.02251$ & $0.02254 $ & $0.00013 $ &  \quad$0.1\sigma$ & $ 0.3\sigma$ \\
$\omega_{\rm{cdm}}$ & $0.1186 $ & $0.1183 $ & \quad$0.0009 $ &  \quad$-0.1\sigma$ & $-0.5\sigma$ \\
$n_{\rm{s}}$      & $0.9662 $ & $0.9668 $ & $0.0037 $ &  \quad$-0.1\sigma$ & $ 0.1\sigma$ \\
$\ln10^{10}A_{\rm{s}}$ & $3.037 $ & $3.044$ & $0.015 $ &  \quad$-0.2\sigma$ & $ 0.3\sigma$ \\
$H_0\left(\frac{\text{km}}{\text{s}\cdot\text{Mpc}}\right)$ & $69.88$ & $69.65$ & $0.50$ &  \quad$<|0.1\sigma|$ & $ -0.5\sigma$ \\
$\sum m_{\nu}$\,(eV)  & $0.07$ & $0.08$ & $0.05$ & \quad $-0.3\sigma$ & $-0.1\sigma$ \\
$b_0$ & $0.999$ & $0.999$ & $0.012$ & \quad$-0.1\sigma$ & $-0.1\sigma$ \\
$\log_{10} \ceff^2$ & $-0.486$ & $-0.060$ & $0.388$ & \quad$0.4\sigma$ & $1.5\sigma$ \\
$w$ & $-1.07$ & $-1.06$ & $0.02$ & \quad$<|0.1\sigma|$ & $0.4\sigma$ \\
$\epi$ & $0.017$ & $0.024$ & $0.013$ & \quad$0.4\sigma$ & $0.9\sigma$ \\	
\end{tabular}
\begin{tabular}{c c c c c c}
\hline
\multicolumn{6}{c}{ii) Neglecting lensing: $\Delta \chi^2 = 2112$} \\
\hline
Parameter & Mean & Best fit & $\sigma$ & \hspace{-0.52cm} shift: Mean & Best fit \\
\hline
$\omega_{\rm{b}}$ & $0.02244$ & $0.02246 $ & $0.00014 $ &  \quad$-0.4\sigma$ & $-0.3\sigma$ \\
$\omega_{\rm{cdm}}$ & $0.1192$ & $0.1194$ & $0.0010$ &  \quad$0.5\sigma$ & $0.7\sigma$ \\
$n_{\rm{s}}$      & $0.9651$ & $0.9652$ & $0.0039$ &  \quad$-0.3\sigma$ & $-0.3\sigma$ \\
$\ln10^{10}A_{\rm{s}}$ & $ 3.032 $ & $3.028 $ & $ 0.015 $ &  \quad$-0.5\sigma$ & $-0.7\sigma$ \\
$H_0\left(\frac{\text{km}}{\text{s}\cdot\text{Mpc}}\right)$ & $69.42$ & $69.38$ & $0.49$ &  \quad$-0.9\sigma$ & $-\sigma$ \\
$\sum m_{\nu}$\,(eV)  & $0.18$ & $0.18$ & $0.04$ &  \quad$2.2\sigma$ & $2.2\sigma$ \\
$b_0$ & $1.027$ & $1.028$ & $0.011$ & \quad$2.4\sigma$ & $2.5\sigma$\\
$\log_{10} \ceff^2$ & $-0.612$ & $-0.429$ & $0.456$ & \quad$0.1\sigma$ & $0.5\sigma$ \\
$w$ & $-1.06$ & $-1.05$ & $0.02$ & \quad$0.4\sigma$ & $0.9\sigma$ \\
$\epi$ & $0.009$ & $0.005$ & $0.012$ & \quad$-0.2\sigma$ & $-0.5\sigma$ \\	
\hline
\end{tabular}
\caption{Results for survey configuration of 10 top-hat redshift bins. MCMC results for ADE model when considering information from the constraints presented in Sec.~\ref{sec:constraints} for the parameters $\omega_{\rm b}$, $\omega_{\rm cdm}$, $n_{\rm s}$, $\ln 10^{10}A_{\rm s}$, $H_0$, $w$, and $\epi$. We show the mean, the best fit, the standard deviation, and the shift of both mean and best fit with respect to the fiducial values in units of the standard deviation. Because the theoretical spectra are computed less accurately than the observed spectra,  deviations smaller than $0.2\sigma$ are not significant.}
\label{Table:t10}
\end{table}

While we use different survey configurations (i.e., number of redshift bins, galaxy selection functions), we can see that a consistent analysis (gray) also taking into consideration lensing convergence can determine most cosmological parameters in the fiducial model. There are however two exceptions: DE sound speed and neutrino mass scale. Firstly, regarding the DE sound speed we notice its  behaviour agrees with previous findings. Galaxy surveys would not be able to accurately determine $\ceff^2$ if its value is close to the speed of light (see, for instance, Ref. \cite{Sapone:2013wda}) which happens  to be the case in our fiducial model. Secondly, we find that disregarding non-linear information, ongoing and forthcoming galaxy surveys will only be able to put upper bounds to the neutrino mass.

A inconsistent analysis neglecting lensing convergence (red) displays a different situation that might also depend on the survey configuration. First, we note that regardless the survey configuration, the neutrino mass $\sum m_\nu$ and the bias amplitude $b_0$ are wrongly determined (i.e., being off by $2-3\sigma$). We conclude neglecting lensing convergence in analyses might lead to a spurious detection of the neutrino mass scale in the ADE model. A seemingly robust result as a number of analyses regarding different cosmological models find alike conclusions concerning $\sum m_\nu$~\cite{Cardona:2016qxn,Cardona:2022mdq,PhysRevD.97.023537}. Second, we do not find relevant disagreement in the value of the DE equation of state $w$ which we introduce in the analysis as having prior information from other experiments. Third, regardless of the survey configuration we wrongly determine a lower Hubble constant (i.e., $\lesssim \sigma$) than in the fiducial value. We conclude that neglecting lensing convergence in analyses of ongoing and upcoming galaxy surveys cannot significantly alleviate the current tension in $H_0$, when the Hubble constant is introduced as having prior information from other probes. Fourth, also included in the inconsistent analysis with a Gaussian prior, the DE anisotropic stress parameter $\epi$ gets pushed towards lower values than in the fiducial model. This result however depends on the survey configuration, hence lacking robustness. Finally, the DE sound speed, which is not taken into consideration as having prior information, also shows a behaviour depending on the survey configuration. While a 5 Gaussian bins analysis wrongly detects $\ceff^2 \neq 1$, we realise this detection vanishes with more redshift bins or a top-hat selection function.      

Note that Figs.~\ref{Fig:g5}-\ref{Fig:t10} display a degeneracy $b_0$-$\sum m_\nu$. An independent probe constraining the bias amplitude $b_0$ might be important for determining $\sum m_\nu$, a target  parameter for ongoing and upcoming galaxy surveys. 

As expected, some of our results depend on the shape of the galaxy redshift bins as well as  the number of redshift bins in which we split the galaxy survey. In Ref.~\cite{Euclid:2021osj}  authors studied the optimal binning for galaxy clustering and found $~10$ bins to be a good choice. While using a greater number of redshift bins might not make significant differences because the analysis of ongoing and upcoming galaxy surveys will be limited by photo-$z$ precision, it heavily increases the required computational resources. We understand the influence of lensing convergence on the inference of parameters will be bigger for a small number of wider redshift bins. In this case radial correlations are suppressed and the constraining power is mainly due to transverse correlations where lensing convergence plays a relevant part. When the galaxy survey is divided in a big number of thinner redshift bins, the number of modes dominated by density and RSD is increased. This increment of modes does not affect lensing convergence, hence neglecting lensing would not be as important as in the case of small number of wide redshift bins.          

\section{Conclusions}
\label{sec:conclusions}

We are witnessing the coming of new, sophisticated data sets that will require careful modelling of observables if biased cosmological constraints are to be avoided. In this paper we join previous investigations and demonstrate that relativistic effects such as lensing convergence play an important part in the analyses of upcoming galaxy surveys and cannot be neglected any longer. We do so by quantifying the bias brought forth by ignoring these effects, for example we find that the bias due to neglecting the lensing convergence is of the order of $0.5-1\sigma$ on average for the matter density parameters and neutrino masses, and $2-3\,\sigma$ for the anisotropic stress parameters (see for example Table \ref{Table:g5}).

Previous works showed the impact of neglecting lensing convergence on the estimation of important quantities such as non-Gaussianity, momentum transfer in a possible dark matter-dark energy interaction, neutrino mass scale, and dark energy equation of state. Here we have further investigated the subject by using a cosmological model describing dark energy by a constant equation of state, constant sound speed, and non-vanishing anisotropic stress.

We considered a phenomenological model for dark energy anisotropic stress which covers general features found in dark energy and modified gravity models and is therefore enough for our purpose. In particular, we regarded a dark energy anisotropic stress sourced by matter perturbations which could emerge, for instance, from coupled dark energy models or DGP-like theories. 

We implemented our anisotropic dark energy cosmological model in a Boltzmann solver and solved the full system of differential equations governing background and linear order perturbations. Then we computed cosmological constraints in our relatively simple extension of the standard cosmological model and found results in good agreement with the Planck Collaboration when using similar data sets. Our results (see Fig.~\ref{Fig:constraints-MI} and Table~\ref{Table:constraints-MI}) do not show any improvement in the discrepancies of $H_0$ and $S_8$ present in the standard cosmological model $\Lambda$CDM. With regard to the dark energy parameters in the model, we find an equation of state $w$ compatible with a cosmological constant value; the dark energy sound speed squared is unconstrained, but close to the speed of light which we set as our upper bound; the dark energy anisotropic stress sourced by matter perturbations is tightly constrained, consistent with a vanishing value, and mainly driven by the inclusion of Redshift-Space-Distortions (RSD) in the data set.     

Finally, we carried out forecasts for the performance of an Euclid-like galaxy survey, also taking into account information from our baseline constraints (i.e., including cosmic microwave background anisotropies, baryon acoustic oscillations, supernovae type Ia, local measurement of the Hubble constant, and RSD) as a Gaussian prior. In the analysis we considered different galaxy survey configurations (i.e., number of redshift bins 5 and 10, as well as top-hat and Gaussian galaxy selection functions). 

Regarding the decrease of uncertainties by including information from the galaxy survey, our findings indicate just a marginal improvement with respect to our baseline constraint. While the estimation of the neutrino mass scale is greatly improved, our analysis reveals that accurate information from non-linear scales might be needed for a detection of $\sum m_\nu$. Here however we were cautious and disregarded non-linear scales when modelling number count fluctuations.   

We found that a consistent analysis including lensing convergence properly determine the values of the fiducial model regardless of the galaxy survey configuration. Nevertheless, the situation is quite different when lensing convergence is neglected in the analysis. First, the dark energy anisotropic stress parameter $\epi$, the dark energy sound speed $\ceff^2$, and the Hubble constant $H_0$ are pushed towards lower values than the fiducial model. These results for dark energy parameters $\epi$ and $\ceff^2$ are however dependent on the survey configuration and lose significance for a 10 top-hat configuration. Although the Hubble constant is consistently determined lower than the fiducial value for all survey configurations, the shift is not big enough to be of any relevance in the current $H_0$ tension. Second, regardless of the survey configuration, the approximation of neglecting lensing convergence induces a heavily biased constraint for the galaxy bias $b_0$ and a spurious detection of the neutrino mass $\sum m_\nu$ (see Figs.~\ref{Fig:g5}-\ref{Fig:t10} and Tables~\ref{Table:g5}-\ref{Table:t10}). Our result for $\sum m_\nu$ in the framework of the anisotropic dark energy model aligns with previous works~\cite{Cardona:2016qxn,Cardona:2022mdq,PhysRevD.97.023537} considering different cosmological models. Therefore, it becomes clear that in order to avoid wrong parameter estimation, correctly modelling galaxy number counts fluctuations (i.e., taking into account lensing convergence) will be necessary in the analysis of ongoing and forthcoming galaxy surveys. 
Another general relativistic effect that could be important in the estimation of the DE sound speed or the neutrino mass is the Doppler effect as was shown in Ref.~\cite{barrera2020relativistic}.

\section*{Acknowledgments}

This work is supported by the Spanish Research Agency (Agencia Estatal de Investigación) through the grant IFT Centro de Excelencia Severo Ochoa SEV-2016-0597 and the Research Projects FPA2015-68048-03-3P [MINECO-FEDER] and PGC2018-094773-B-C32. W.C. acknowledges financial support from the Departamento Administrativo de Ciencia, Tecnología e Innovación (COLCIENCIAS) under the project `Discriminación de modelos de energía oscura y gravedad modificada con futuros datos de cartografiado galáctico' and from Universidad del Valle under the contract 449-2019. S.N. acknowledges support from the Ram\'{o}n y Cajal program through Grant No. RYC-2014-15843. This research was supported by resources supplied by the Hydra cluster in the Institute of Theoretical Physics IFT CSIC/UAM. The statistical analyses as well as the plots were made with the Python package GetDist.\footnote{\url{https://github.com/cmbant/getdist}}

\section*{Numerical codes}

Modified \texttt{CLASS} code reproducing results in this work can be found in the GitHub branch \texttt{LDE} of the repository  \href{https://github.com/wilmarcardonac/EFCLASS.git}{\texttt{EFCLASS}}. 

\section*{Appendix: Specifications of the Euclid-like photometric survey}
\label{section:appendix}

Having the primordial power spectrum of curvature perturbations 
$\mathcal{P}_{\mathcal{R}}(k)=A_s k^{n_s-1}$ and using the transfer function $\Delta_{\ell}^i(k)$ at redshift $z_i$, we can write the angular power spectra 
\begin{equation} \label{eq:Cl}
C_{\ell}^{ij} = 4\pi \int d\ln k\; \mathcal{P}_{\mathcal{R}}(k) \Delta_{\ell}^{i}(k) \Delta_{\ell}^{j}(k) \;,
\end{equation}
that is, as an integral of the product of transfer functions over wave numbers $k$.  
The transfer function $\Delta_{\ell}^i(k)$ is actually computed as 
\begin{equation} \label{eq:Delta_l}
  \Delta_{\ell}^i(k) = \int dz\; \frac{dN}{dz} W_i(z) \Delta_{\ell}(z,k) \;,
\end{equation}
where the window function $W_i(z)$ describes the binning in redshift, and $dN/dz$ is the number of galaxies per redshift interval.

The transfer functions $\Delta_{\ell}(z,k)$ in Eq.~\eqref{eq:Delta_l} read 
\begin{eqnarray}
  \Delta_{\ell}(z,k) &=& b(z) \delta(z,k) j_\ell(kr(z))
  + \frac{k}{\cal H} \tilde{V}(z,k) \frac{d^2 j_\ell(kr(z))}{d(kr(z))^2}
  \nonumber \\
  &&+ \left( \frac{2-5s}{2}\right) \ell(\ell+1)
  \nonumber \\
  &&\times \int_0^{r(z)} d\tilde r\; \frac{r(z)-{\tilde r}}{r(z) {\tilde r}} \left[ {\Phi}(\tilde z,k) + {\Psi}(\tilde z,k) \right] j_\ell(k{\tilde r}) \;,
  \nonumber \\
  \label{Eq:tf}
\end{eqnarray}
where the first term is the intrinsic galaxy density perturbation, the second term corresponds to redshift space distortions, the last term denotes lensing convergence effects, and  $j_\ell(kr(z))$ are spherical Bessel functions. In Eq.~\eqref{Eq:tf} we use the Fourier transforms of: i) velocity potential $v_i\equiv-\partial_i \tilde{V}$ (in the Newtonian gauge and with initial conditions ${\cal R}(z_{\rm{in}},k)=1$); ii) gravitational potentials $\Phi$ and $\Psi$; iii) density perturbations in the comoving gauge.    

This work focuses on the last term in Eq.~\eqref{Eq:tf}, namely, the integral along the line of sight. Lensing convergence plays a part in number counts by magnifying the sources, thus leading to changes in their number density per steradian. For a given galaxy population, the magnification bias $s(z)$ depends on its luminosity function. 

As in Refs. \cite{Laureijs:2011gra,Amendola:2016saw}, we regard the Euclid-like photometric survey as having a number of galaxies per redshift and per steradian
\be
\frac{dN}{dzd\Omega} = 3.5\times10^8 z^2 \exp\left[-\left( \frac{z}{z_0} \right)^{3/2}\right] \; \quad \mbox{for} \quad 0<z<2.0\;, \label{EQ:dndzdo} \\
\ee
galaxy density 
\be
d=30\mbox{ arcmin}^{-2}\;, \label{Eq:d}\\
\ee
covered sky fraction
\be 
f_{\rm sky}=0.364\;,\\
\ee
galaxy bias as given in Eq.~\eqref{Eq:bias},
and magnification bias
\be 
s(z)=s_0 + s_1 z + s_2 z^2 + s_3 z^3 \;. \label{eq:sz_euclid}
\ee
In Eqs.~\eqref{EQ:dndzdo}-\eqref{eq:sz_euclid}
$z_0=z_{\rm mean}/1.412$ and the median redshift is $z_{\rm mean}=0.9$. Coefficients in Eq.~\eqref{eq:sz_euclid} are $s_0=0.1194$, $s_1=0.2122$, $s_2=-0.0671$, and $s_3=0.1031$ as computed in
 Ref.~\cite{Montanari:2015rga}.

In our computations we set the lower redshift bound to $z=0.1$ and assume that within each redshift bin both galaxy bias and magnification bias are determined by the mean redshift of the bin.

\bibliographystyle{elsarticle-num}
\bibliography{biblio}

\end{document}